\PassOptionsToPackage{
	breaklinks=true,
	colorlinks=true,
}{hyperref}

\documentclass[preprint,5p,sort&compress,lefttitle]{elsarticle}

\usepackage[british]{babel}
\usepackage[T1]{fontenc}
\usepackage[utf8]{inputenc}  

\usepackage{amsmath,amsfonts,amssymb,amsthm}
\usepackage{booktabs,multirow,relsize,subcaption,graphicx,xifthen,xspace,xfrac}
\usepackage[dvipsnames,table]{xcolor}
\usepackage{tikz,marginnote}
\usepackage{threeparttable}  
\usepackage[inline]{enumitem}
\usepackage[pagewise]{lineno}  
\usepackage[hyphens]{url}  
\usepackage[normalem]{ulem}
\usepackage[edges]{forest}
\usepackage{breakcites}
\usepackage{ragged2e}  
\usepackage{orcidlink}
\usepackage{marginnote,tokcycle}

\RequirePackage{hyperref}
\usepackage[nameinlink]{cleveref}

\graphicspath{{figures/},}  

\newboolean{tosubmit}
\setboolean{tosubmit}{true}

\newboolean{showlinums}
\setboolean{showlinums}{false}

\ifthenelse{\boolean{showlinums}}{%
  \usepackage[pagewise]{lineno}  
  \linenumbers

}{}

\ifthenelse{\boolean{tosubmit}}{}{ 
  \usepackage[firstpage]{draftwatermark}  
  \SetWatermarkColor{Blue!11}
}

\makeatletter  
\renewcommand{\elsparagraph}{\@startsection{paragraph}{4}{0\z@}%
  {7\p@ \@plus 3\p@ \@minus 2\p@}%
  {-6\p@}%
  {\normalfont\bfseries}}
\makeatother

\Crefname{equation}{eq.}{eqs.}
\crefname{equation}{equation}{equations}
\Crefname{figure}{Fig.}{Figs.}
\crefname{figure}{figure}{figures}
\Crefname{tabular}{Table}{Tables}
\crefname{tabular}{table}{tables}
\Crefname{definition}{Def.}{Defs.}
\crefname{definition}{definition}{definitions}
\Crefname{proposition}{Prop.}{Props.}
\crefname{proposition}{proposition}{propositions}
\Crefname{section}{Sec.}{Secs.}
\crefname{section}{section}{sections}

\makeatletter
\def\@firstoftwo@third#1#2#3#4#5{%
  \def\temp##1.##2\@nil{##2}%
   \temp#1\@nil}
\newcommand\sref[1]{%
   \expandafter\@setref\csname r@#1\endcsname\@firstoftwo@third{#1}%
}
\makeatother

\makeatletter
\def\THICKhrulefill{\leavevmode \leaders \hrule height 5pt\hfill \kern \z@}
\def\getfirst#1#2\relax{\tctestifnum{\count@stringtoks{#1}>1}{ERROR}{#1}}
\makeatother
\newcommand{\colorpar}[3]{\colorbox{#1}{\parbox{#2}{#3}}}
\newcommand{\marginremark}[3]{%
  \ifthenelse{\boolean{tosubmit}}{}{
	\marginnote{\raggedrightmarginnote\colorpar{#2}{.8\linewidth}%
      {\raggedrightmarginnote\color{#1}#3}}
}}
\newcommand{\textremark}[5]{%
  \ifthenelse{\boolean{tosubmit}}{}{
  \marginremark{#1}{#2}{\tiny\sffamily{[#3]\ #5}}%
  {\def\ULthickness{.8pt}\color{#1!80!black}\uline{#4}}
}}
\newcommand{\highlightedremark}[4]{%
  \ifthenelse{\boolean{tosubmit}}{}{
	\begin{center}\fcolorbox{#1}{#2}{%
	\begin{minipage}{.98\linewidth}\color{#1}%
	\textbf{\THICKhrulefill[ #3 ]\THICKhrulefill}%
	\par\noindent#4\end{minipage}}\end{center}%
}}
\newcommand{\hey}[4]{%
  \ifthenelse{\boolean{tosubmit}}{}{
  \reversemarginpar
  \leavevmode\marginnote{\sffamily\Large\color{#1}@\getfirst#3\relax\relax}
  \colorbox{#2}{\sffamily\bfseries{@#3:}}~{\sffamily\color{#1}#4}}
  \normalmarginpar}
\newcommand{\TODO}[1]{%
  \ifthenelse{\boolean{tosubmit}}{}{
  \noindent\textsf{\color{Red}\textbf{TODO:} #1}%
  \marginnote{\textsf{\color{red}\bfseries TODO}}}}
\newcommand{\tocite}[1][??]{%
  \ifthenelse{\boolean{tosubmit}}{}{
  \noindent\textbf{\sffamily\textcolor{blue!85}{[#1]}}%
  \marginnote{\textsf{\color{blue}\bfseries CITE!}}}}
\colorlet{FM-fg}{Plum}
\colorlet{FM-bg}{Tan!12}
\colorlet{YF-fg}{BrickRed}
\colorlet{YF-bg}{orange!11}
\colorlet{CEB-fg}{TealBlue!70!black}
\colorlet{CEB-bg}{Aquamarine!8}
\colorlet{NNN-fg}{WildStrawberry!75!black}
\colorlet{NNN-bg}{Peach!33}

\newcommand{\hrmkCEB}[1]{\highlightedremark{CEB-fg}{CEB-bg}{CEB}{#1}}

\colorlet{shitbrown}{Brown!30!Maroon!90!black!90}
\NewDocumentEnvironment{shit}{ O{Fabio: ignore this!} }
  {\bigskip\begingroup\color{shitbrown}\slshape\smaller[1]}
  {\endgroup\medskip}

\iffalse  
  \definecolor{darkred}{rgb}{0.8,0,0}
  \definecolor{darkgreen}{rgb}{0.0,0.55,0.05}
  \newcommand{\reviewnote}[1]{\marginnote{\footnotesize{#1}}\ignorespaces}
\else

  \newcommand{\reviewnote}[1]{}
\fi
\newcounter{reviewer}
\crefname{reviewer}{$R$}{$R$}
\Crefname{reviewer}{Reviewer}{Reviewers}
\crefformat{reviewer}{#2\textit{R#1}#3}
\Crefformat{reviewer}{#2Reviewer\:#1#3}
\newcounter{comment}[reviewer]
\makeatletter  
  \renewcommand{\p@comment}{\p@reviewer\thereviewer.}
\makeatother
\crefname{comment}{$R$}{$R$}
\Crefname{comment}{Comment}{Comments}
\crefformat{comment}{#2\textit{R#1}#3}

\NewDocumentEnvironment{reply}{}
  {\par\vspace{1ex}\noindent\begingroup\itshape
   \textsf{Reply:}\ignorespaces~\begin{minipage}[t]{.9\linewidth}\raggedright}
  {\end{minipage}\endgroup}

\usetikzlibrary{positioning,fit,patterns,patterns.meta}

\newcolumntype{L}[1]{>{\RaggedRight\let\newline\\\arraybackslash}m{#1}}
\newcolumntype{C}[1]{>{\Centering\let\newline\\\arraybackslash}m{#1}}
\newcolumntype{R}[1]{>{\RaggedLeft\let\newline\\\arraybackslash}m{#1}}

\colorlet{shade1}{black!8}
\colorlet{shade2}{white}
\def\colortablepreamble{%
  \renewcommand{\arraystretch}{1.06}
  \setlength{\tabcolsep}{2.45pt}
  \setlength{\aboverulesep}{-.5pt}
  \setlength{\belowrulesep}{.25pt}
  \setlength{\extrarowheight}{.45ex}
  \newcolumntype{L}[1]{>{\raggedright\arraybackslash}m{##1}}
  \newcolumntype{C}[1]{>{\centering\arraybackslash}m{##1}}
  \newcolumntype{R}[1]{>{\raggedleft\arraybackslash}m{##1}}
  \rowcolors{2}{shade1}{shade2}
}

\newlist{fileslist}{enumerate}{1}
\setlist[fileslist]{
	start      = 1,
	topsep     = .5ex,
	parsep     = 0pt,
	leftmargin = 2em,
	label      = (\parbox{.6em}{\centering\textit{\alph*}}),
}
\Crefname{fileslisti}{File}{Files}

\newlist{columnslist}{enumerate}{1}
\setlist[columnslist]{
	start      = 1,
	topsep     = .5ex,
	parsep     = 0pt,
	leftmargin = 3.3em,
	label      = \raisebox{1pt}{\parbox{2.5em}{col.\:\textrm{\Alph*}}},
	ref        = \textrm{\Alph*},
}
\Crefname{columnslisti}{Col.\!}{Columns}

\definecolor{folderbg}{RGB}{124,166,198}
\definecolor{folderborder}{RGB}{110,144,169}
\def\mySize{4pt}
\tikzset{
  folder/.pic={
    \filldraw[draw=folderborder,top color=folderbg!50,bottom color=folderbg]
      (-1.05*\mySize,0.2\mySize+5pt) rectangle ++(.75*\mySize,-0.2\mySize-5pt);
    \filldraw[draw=folderborder,top color=folderbg!50,bottom color=folderbg]
      (-1.15*\mySize,-\mySize) rectangle (1.15*\mySize,\mySize);
  }
}
\forestset{
  is file/.style={edge path'/.expanded={%
    (!u.south west) +(7.5pt,0) |- ([xshift=-2pt].child anchor)},
    inner sep=1pt
  },
  this folder size/.style={edge path'/.expanded={%
    (!u.south west) +(7.5pt,0) |- (.child anchor) pic[solid]{folder=#1}},
    inner ysep=0.6*#1
  },
}

\DeclareMathOperator{\conflict}{cfl}
\DeclareMathOperator{\misses}{misses}
\DeclareMathOperator{\matches}{matches}
\DeclareMathOperator{\choices}{choices}

\newcommand{\code}[1]{\ensuremath{\mathtt{\smaller#1}}\xspace}
\newcommand{\acronym}[1]{\ensuremath{\textsc{\larger{#1}}}\xspace}
\newcommand{\card}[1]{\ensuremath{\left\vert{#1}\right\vert}\xspace}
\newcommand{\tuple}[1]{\ensuremath{\left\langle{#1}\right\rangle}\xspace}
\newcommand{\Cpp}{C\raisebox{.3pt}{\kern-.4pt+\kern-.8pt+}\xspace}
\newcommand{\NVD}{\acronym{nvd}}             
\newcommand{\CVE}{\acronym{cve}}             
\newcommand{\CVEs}{\CVE{s}\xspace}
\newcommand{\PyPI}{\acronym{p}{y}\acronym{pi}}  
\newcommand{\PEP}{\acronym{pep}}             
\newcommand{\FOSS}{\acronym{foss}}           
\newcommand{\ML}{\acronym{ml}}               
\newcommand{\NLP}{\acronym{nlp}}             
\newcommand{\SE}{\acronym{se}}               
\newcommand{\ttg}{\code{g}}
\newcommand{\tta}{\code{a}}
\newcommand{\ttv}{\code{v}}
\newcommand{\ga}{\code{\ttg{:}\tta}}
\newcommand{\gav}{\code{\ttg{:}\tta{:}\ttv}}
\newcommand{\CVSS}{\acronym{cvss}}
\newcommand{\ASS}{\ensuremath{\mathtt{ASS}}\xspace}

\long\protect\def\explanation#1{\begin{minipage}{0.9\linewidth}\small #1\end{minipage}}

\def\TITLESHORT{Cross-ecosystem categorization}
\def\TITLELONG{A manual-curation protocol for the categorization of Java Maven libraries along Python PyPI Topics}
\hypersetup{%
    pdftitle={\TITLELONG},
    pdfauthor={Carlos E. Budde, Ranindya Paramitha, Yuan Feng, Fabio Massacci},
    pdfkeywords={Library vulnerability, cybersecurity},
}

\begin{document}

\title{\TITLESHORT: \TITLELONG}

\author[1]{Ranindya Paramitha\,\orcidlink{0000-0002-6682-4243}\,\corref{cor1}}
	\ead{ranindya.paramitha@unitn.it}
\author[1]{Yuan Feng\,\orcidlink{0000-0001-5401-8597}\,}
	\ead{yuan.feng@unitn.it}
\author[1,2]{Fabio Massacci\,\orcidlink{0000-0002-1091-8486}\,}
	\ead{f.massacci@ieee.org}
\author[1]{Carlos E.\ Budde\,\orcidlink{0000-0001-8807-1548}\,\corref{cor1}}
	\ead{carlosesteban.budde@unitn.it}
\address[1]{%
	University of Trento,
	Trento,
	I-38122,
	Italy.
}
\address[2]{%
	Vrije Universiteit Amsterdam,
	Amsterdam,
	1081\;HV,
	The~Netherlands.
}
\tnotetext[tn1]{This work was funded by the EU under GAs
	101067199 (\emph{Pro\-SVED}),
	101120393 (\emph{Sec4AI4Sec}), and
	952647 (\href{https://assuremoss.eu/en/}{\emph{H2020-AssureMOSS}}).
}
\cortext[cor1]{Corresponding authors.}

\begin{abstract}
\textbf{Context:}
Software of different functional categories, such as text processing vs.\ networking, has different profiles in terms of metrics like security and updates.
Using popularity to compare e.g.\ Java vs.\ Python libraries might give a skewed perspective, as the categories of the most popular software vary from one ecosystem to the next.
How can one compare libraries datasets across software ecosystems, when not even the category names are uniform among them?
\textbf{Objective:}
We study how to generate a language-agnostic categorisation of software by functional purpose, that enables cross-ecosystem studies of libraries datasets.
This provides the functional fingerprint information needed for software metrics comparisons.
\textbf{Method:}
We designed and implemented a human-guided protocol to categorise libraries from software ecosystems.
Category names mirror PyPI Topic classifiers, but the protocol is generic and can be applied to any ecosystem.
We demonstrate it by categorising 256 Java/Maven libraries with severe security vulnerabilities.
\textbf{Results:}
The protocol allows three or more people to categorise any number of libraries.
The categorisation produced is functional-oriented and language-agnostic.
The Java/Maven dataset demonstration resulted in a majority of Internet-oriented libraries, coherent with its selection by severe vulnerabilities.
To allow replication and updates, we make the dataset and the protocol individual steps available as open data.
\textbf{Conclusions:}
Libraries categorisation by functional purpose is feasible with our protocol, which produced the fingerprint of a 256-libraries Java dataset.
While this was labour intensive, humans excel in the required inference tasks, so full automation of the process is not envisioned.
However, results can provide the ground truth needed for machine learning in large-scale cross-ecosystem empirical studies.
\end{abstract}


\maketitle

\section{Introduction}
\label{sec:intro}

Java and Python are among the most-used programming languages in the world---a trend expected to continue for the foreseeable future \cite{PopLangsRankingGitHub,PopLangsRanking}.
Chief reasons behind their popularity are their cornucopia of libraries and the existence of a central repository, all of which creates \emph{ecosystems}
that provide official sources for them.
Here we use the term \emph{library} broadly, to refer to any software package, project, or functionality identified by a top-level URL in a software repository that defines an ecosystem.

Software ecosystems have been the object of intensive research in Software Engineering (\SE).
However, while \SE studies are common for ecosystems taken individually, cross-ecosystem studies are relatively rare.
For instance, the past two years of the MSR track on \textsl{Data/Tool Showcase} report 13 studies on mining Python repositories, and 17 on mining Java repositories, but only 4 on both. 

Moreover, most studies are cross-language as opposed to cross-ecosystem \cite{NRRS20,NAR+22}.
Cross-eco\-sys\-tem studies require common library categories for classifications~\cite{BYJ19,DMG19,GHZ23,XWC+23}.
For example, one can expect that libraries providing Internet-facing functionalities will be more subject to scrutiny for software vulnerabilities, in comparison to libraries used for syntax highlighting.
In view of this, when an ecosystem contains a disproportionate number of libraries of the latter type (e.g.\ because they are the most popular), one might interpret that it is less prone to security vulnerabilities.
And yet the opposite might be true, once representative functionalities are considered---but this necessitates a taxonomy of library functionalities applicable to the ecosystems under comparisons.

\begin{figure*}  
	\centering
	\begin{subfigure}{.47\linewidth}
		\centering
		\includegraphics[width=\linewidth]{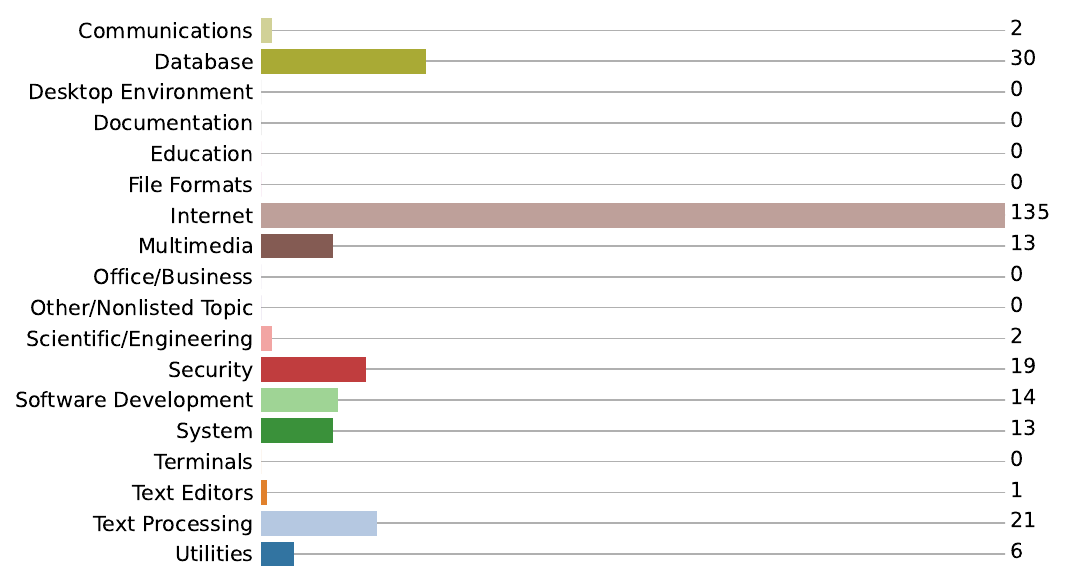}
		\caption{256 Java/Maven libraries with high or critical vulnerabilities
		         \cite{PFMB24}}
		\label{fig:barschart:Maven}
	\end{subfigure}
	\hfill
	\begin{subfigure}{.47\linewidth}
		\centering
		\includegraphics[width=\linewidth]{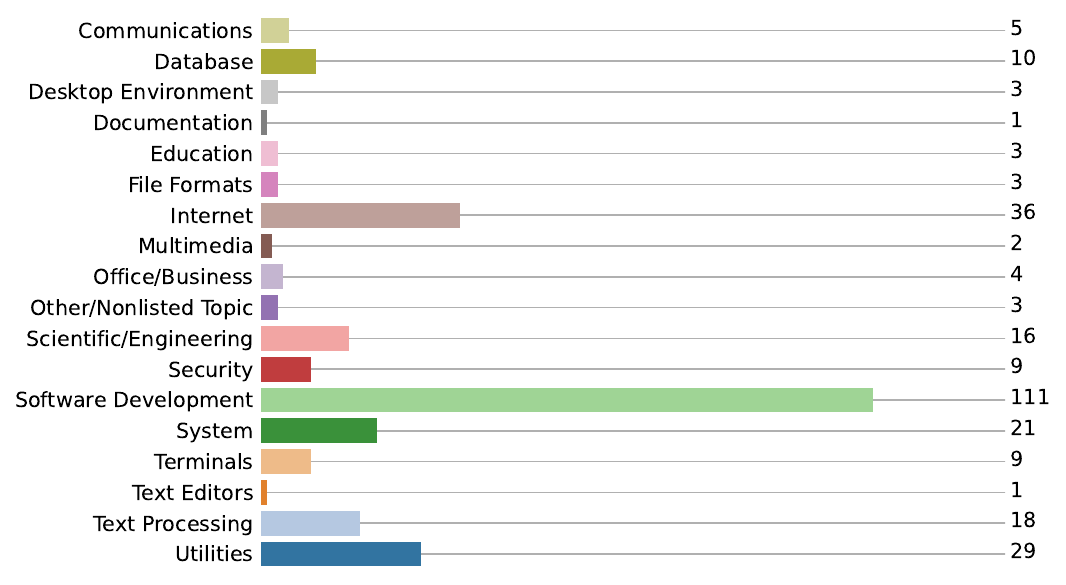}
		\caption{Top-256 popular Python/PyPI libraries \cite{KTSD23}}
		\label{fig:barschart:PyPI}
	\end{subfigure}
	\explanation{%
	\medskip%
	Cross-ecosystem studies must select functionally-equivalent sets of libraries, or measure the differences, lest situations like this inadvertently bias results:
	While popular Python libraries are skewed towards \code{Software} \code{Development}, Java libraries with high or critical vulnerabilities are skewed towards \code{Internet};
	A proper case-control study could conclude that \textit{``Java is more prone to vulnerabilities than Python''} only if both distributions are similar \cite{AM14,VWM16}, e.g.\ a Python sample with an \code{Internet} to \code{Software} \code{Dev.} ratio around \sfrac{135}{14} ---but that is the opposite of the situation above.
	}
	\caption{Distribution of libraries along \PyPI Topics in two ecosystems: Java/Maven and Python/\PyPI}
	\label{fig:barschart}
\end{figure*}

\smallskip
Our long-term aim is to compare security aspects across the Java/Maven and Python/\PyPI ecosystems.
Such comparisons can only be performed under the light of cross-ecosystem library taxonomies.
More specifically, a proper con\-trol-\-case study requires
\begin{enumerate*}[label=(\textit{\alph*}),itemjoin={,},itemjoin*={{, and }}]
\item	using the same categorisation for all the libraries---in
		both ecosystems---that will be part of the study
\item   filtering libraries based on these categories
		in order to ensure comparable datasets
\end{enumerate*}
\cite{AM14,VWM16}.

The importance of this is illustrated by \Cref{fig:barschart}, which shows how popular Python libraries are skewed towards \code{Software} \code{Development}, while Java libraries with high or critical vulnerabilities are skewed towards \code{Internet} \cite{KTSD23,PFMB24}.
Any comparison between these two datasets should account for such disparity \cite{AM14}; or better still, the datasets should be composed of libraries with a similar functional fingerprint---unlike the situation in \Cref{fig:barschart:Maven} vs.\ \Cref{fig:barschart:PyPI}.

\subsection{Objective and contributions}
\label{sec:intro:contributions}

Given the limitations of Maven categories and tags to provide an overarching coherent classification of libraries functionality---see \Cref{sec:motivation}---and the taxonomy available as \PyPI Topics better suited for that purpose, \emph{we designed a protocol to label libraries from any ecosystem with PyPI Topics, and demonstrate its application to the Java/Maven ecosystem.}
More in detail, our contributions are:

\begin{itemize}[leftmargin=1.5em,parsep=0pt,topsep=.5ex]
\item
\emph{The definition of a protocol to assign the 24 highest-level PyPI Topic classifiers to Java libraries},
interpreting them as the names of functional categories.
	\begin{itemize}[leftmargin=1.5em]
	\item Our protocol introduces natural-language descriptions of the functions covered by each \PyPI Topic, together with guidelines, application examples, and rules of thumb; it also details the roles and number of people needed at each stage.
	\item The protocol can be used to replicate the study, or perform cross-ecosystem comparisons from Python/ \PyPI to other ecosystems such as JS/npm. 
	\end{itemize}
\item
\emph{The selection of 256 Java libraries from Maven Central},
filtering them by code availability (we worked with \FOSS), and the presence of a \underline{C}ommon \underline{V}ulnerability and \underline{E}xposure entry (\CVE) affecting them with a high or critical \underline{C}ommon \underline{V}ulnerability \underline{S}coring \underline{S}ystem value, i.e.\ libraries with a \CVE with $\CVSS \geqslant 7$.
\item
\emph{The execution of our protocol for the selected Maven libraries},
registering all material consulted, as well as the information produced at each stage.
	\begin{itemize}
	\item \emph{A dataset with the results is made publicly available online} \cite{PFMB24} for further \SE research. 
	\end{itemize}
\end{itemize}

\paragraph{Outline}
After 
discussing related work in \Cref{sec:related_work}, we introduce in \Cref{sec:protocol} a protocol to categorise libraries from any ecosystem along \PyPI Topics.
Then, \Cref{sec:results} describes the results obtained by applying that protocol to 256 libraries from the Java/Maven ecosystem.
We conclude in \Cref{sec:conclu} with some foreseen applications, as well as known limitations and possible extensions to our work.

\paragraph{Data and Artefacts Availability}
We provide online public access to all files and programs generated and used in this work, to make them available for \SE studies and in particular cross-repository studies, in \cite{PFMB24} and under the following link:
\href{https://zenodo.org/records/10480832}{\texttt{\color{blue}https://zenodo.org/records/10480832}}.

\section{Motivating examples}
\label{sec:motivation}

The Python programming language implements centralised enhancement proposals known as \PEP{s}.  
\PEP{s} regulate packages metadata in the leading Python ecosystem known as the Python Package Index (\PyPI).
\PEP~301%
\footnote{%
	\url{https://peps.python.org/pep-0301/\#distutils-trove-classification}%
}
introduced Trove classifiers that include a \emph{Topic} field, with 24 possible categories that describe the functional domain of a package.
In particular, Trove classifiers---of which \PyPI Topics are an example---are designed for archiving, which yields the classical taxonomies used for hu\-man{-}\-de\-veloped documents and libraries.
In other words, \PyPI Topics can serve as functional categories for libraries.

Also Maven Central---the leading Java ecosystem---has an extensive categorisation into over 150 groups \cite{Maven}.
The categories of a library in Maven are defined by its developers, manually or semi-automatically \cite{MavenProjectDescriptor,MavenRepositoryTags}.
This has resulted in categories as broad as \code{I/O} \code{Utilities},
and as narrow as \code{Atlassian} \code{Plugins}.
In general, organic categorisations like Maven's can yield groups of software projects with similar characteristics from the perspective of the developers of a library, but not necessarily from that of its users, or for \SE research.
More importantly, such categorisations lack an overarching coherent structure that hinders cross-ecosystem studies, as we illustrate next.

\smallskip
To see how library categories impact \SE research, consider our goal of studying security aspects of Java/Maven libraries for comparison against Python/\PyPI ones.
Since there is no ``\code{Security}'' category in Maven, the first question is how to gather all Java libraries relevant to our study.

An option would be to work with all categories subsumed by our (user-sided) expectation of what \code{Security} means.
This could include Maven categories such as \code{Auth} \code{Libraries}, \code{Security} \code{Frameworks}, etc.

An issue is that no Maven category has a description besides its textual name, e.g.\ no sub-categories are listed.  
Therefore, the question of which categories to include \emph{must be decided by the researchers ad hoc, for the specific work at hand}.
For us, the lack of guidelines and descriptions of what are Java libraries with security functionality hinders reproducibility; in general, it also complicates the extension of scientific works across ecosystems.

Another issue is what to do with Maven categories that are ambiguous with respect to the selected functionality.
Consider e.g.\ \code{Jenkins} \code{Plugins}: while most libraries in this category are irrelevant for security, some implement e.g.\ authentication and must be included.
Maven tags could help to separate wheat from chaff here, but unfortunately they are insufficient.
For instance, the artifact \code{org.jenkins\text{-}ci.plugins:} \code{authorize\text{-}project} (used to \textsl{``Configure projects to run with specified authorization''}) falls within scope, but it would be missed because its tags are \code{plugin}, \code{build}, and \code{jenkins}.

The above illustrates yet another issue with organic classifications: they originate from a small group of people with a potentially different perspective of the software than its users.
As another example, \code{com.clever\text{-}cloud:} \code{biscuit\text{-}java} is described in Maven as \textsl{``Biscuit authentication Java library''}.
However it has no category, despite the existence of the category \code{Auth} \code{Libraries}, and it is only tagged as \code{cloud}.
The absence of category and tags makes it impossible to capture such libraries, causing gaps in (security) \SE studies in the Java/Maven ecosystem.

\smallskip
The situation improves when using overarching categorisations such as \PyPI Topic classifiers, especially when garnished with explicit rules and example that help humans to understand what is the functional coverage of each category.
While subjectivity can never be ruled out completely, \emph{taxonomies like \PyPI Topics yield better functional divisions than organic, ecosystem-specific categorisations, ultimately resulting in less fragmentation and redundancy.}
All of this benefits research transparency, and enables case-control studies across ecosystems.

\section{Scope and related work}
\label{sec:related_work}

To the best of our knowledge, this is the first time that a protocol is devised and applied for a human, systematic, cross-ecosystem categorisation of \FOSS libraries.
By \emph{library} we denote a \ga coordinate that singles out a group and artifact in Maven Central.
This is generalisable to the language-agnostic concept of code---e.g.\ a software project, package, or low-level program---identified by a top-level URL in a public repository that defines an ecosystem.

Our work differs from \SE initiatives like \emph{Awesome Java}, which strives to increase developers' performance by providing a search engine, clickable hyperlinks, and project-wise classifications \cite{AwesomeJava}.
In contrast, our focus is scientific: we designed a protocol to assign categories from a func\-tion\-al\-ity-\-oriented taxonomy---\PyPI Topic classifiers---to libraries written in any language, with the aim to bootstrap the study of relationships between these functionalities and security issues across different ecosystems.

\paragraph{Cross-ecosystem studies}
Similar works exist that compare libraries across software ecosystems, e.g.\ Java/Maven, Ruby/gems, JS/npm, etc. \cite{BYJ19,DMG19,GHZ23,XWC+23}.
To avoid processing the full source code of each library, these works rely on \SE metrics to cluster libraries together: the categorisation that our protocol produces is an important metric for such clusterisation.
For instance, \cite{XWC+23} studies cross-ecosystem vulnerabilities, and \cite{DMG19} compares the evolution of library dependencies in various environments.
In both cases it is important to distinguish libraries that are highly exposed to the Internet (such as remote filesystem implementations), from those that are not (such as text editors).
Because the higher exposition to cyberattacks of Internet-oriented libraries fosters the discovery of more vulnerabilities, causing more frequent software updates.
Thus, selecting libraries e.g.\ by popularity alone will yield results biased by the functionalities of the popular libraries in the ecosystem, as illustrated in \Cref{fig:barschart}.
Our protocol helps to measure and correct that bias, by producing uniform functional labels for the libraries in any ecosystem.

\paragraph{Datasets with categorised libraries}
Works like \cite{LCC22,DSB21,PSZ19,CGZ17,BHV16,RDV13} focus on the annotated dataset of Java libraries that they present.
From these, two stand close to our categorisation objectives:
\begin{itemize*}[itemjoin={{; }}]
\item[\cite{LCC22}]
employed a semi-automatic approach---a user-guided wizard---to categorise software projects modelled in the Eclipse Modelling Framework (a well-defined subset of Java libraries), with a focus on the \emph{models} rather than the libraries
\item[\cite{BHV16}]
offers a labelling of Java GitHub projects into six categories, selected based on their \emph{GitHub description}
\end{itemize*}.
In contrast to these works, our protocol is designed to label libraries (from any ecosystem) with one of 24 categories, for which it employs information from several sources per library that describe its functional purpose.
Generalising, our protocol resembles an annotation schema with guidelines over a software-functionality ontology, for which no general annotation tools are reported in recent surveys \cite{NS19}.

\smallskip
Our long-term aim involves the study of relationships between software functions and (public disclosure of) security vulnerabilities.
In that sense we are also close to works like \cite{VD20,VD20b,ASMM18,Esc15,TRP09,KGMI06}, which propose code-driven categorisations of software using bulk-processing tools, such as those typically found in Natural-Lan\-guage Processing (\NLP) and Machine Learning (\ML) applications.

\paragraph{ML related work}
\cite{VD20,VD20b} use a plethora of supervised \ML algorithms to read the source code of Java projects and assign them a Maven category or tags.
The results obtained are at best as informative as Maven categories and tags are, which brings the issues mentioned in \Cref{sec:motivation}.
In contrast, \cite{Esc15} applies unsupervised \ML on bytecode to cluster libraries based on functionality.
Here, too, categories are part of the input data, and the final result can be only as informative as the initial categorisation was.
Instead, our protocol serves to decide which functional category---comparable to \PyPI libraries---suits a library best, augmenting the data available in other ecosystems such as Maven.

\paragraph{NLP related work}
Works like \cite{ASMM18,TRP09,KGMI06} can be more versatile than those mentioned above, specially in cases like \cite{KGMI06} where category names are chosen from the corpus.
In that case, however, changing the corpus changes the result, so cross-corpora comparisons are hindered if not altogether impossible.
In contrast, \cite{ASMM18,TRP09} rely on category names that come with the input data, and use \acronym{lda} to derive most-likely categories for the (commented) source-code of a project.
Here, too, categories are an input, which is the information that our protocol \emph{produces}.
Moreover, while the largest ground truth for demonstration contains 103 pre-labelled libraries \cite{ASMM18}, the final example dataset presented in this work contains 256 categorised libraries \cite{PFMB24}.
Furthermore, we assigned a category to a library after different actors have read four human-written descriptions about the purpose of that library, which is a more accurate and rich source of information---of the functional purpose of the library---than its \NLP-processed source code.

\begin{figure*}
	\centering
	\includegraphics[width=.95\linewidth]{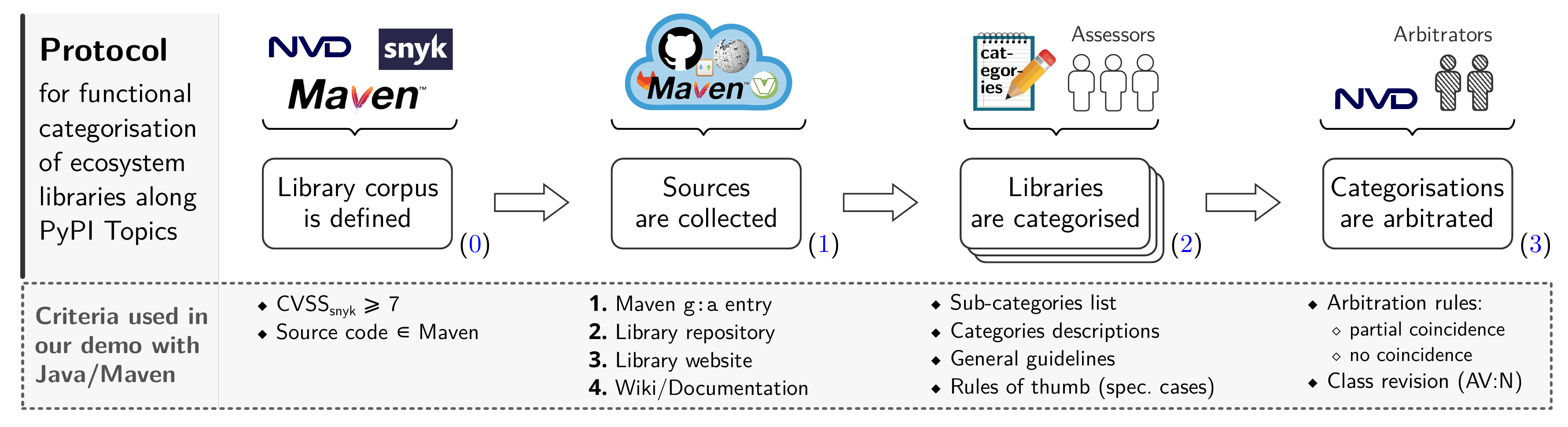}
	\explanation{%
The initial step (\protect\sref{sec:protocol:libraries}) defines the libraries of interest: we use FOSS from Maven with high or critical security vulnerabilities.
To categorise libraries with \PyPI Topics the first step (\protect\sref{sec:protocol:sources}) is to gather the sources.
These provide data about each library, read separately by each assessor to choose categories for it (\protect\sref{sec:protocol:assess}).
Ambiguous assessments are arbitrated~(\protect\sref{sec:protocol:arbitrate}).}
	\caption{High-level description of the protocol, here instantiated for the Java/Maven ecosystem---see \Cref{sec:protocol}}
	\label{fig:protocol}
\end{figure*}

\section{Methodology (categorisation protocol)}
\label{sec:protocol}

Our protocol, synthesised in \Cref{fig:protocol}, differentiates four stages:
\begin{itemize*}[itemjoin={{, }}, itemjoin*={{, and }}]
\item[(\sref{sec:protocol:libraries})] definition of the libraries
\item[(\sref{sec:protocol:sources})]   acquisition of the sources
\item[(\sref{sec:protocol:assess})]    assessment of the libraries
                                          (i.e.\ categorisation)
\item[(\sref{sec:protocol:arbitrate})] arbitration of the assessments.
\end{itemize*}

%
\begin{table}
  \centering
  \caption{Roles of actors in the protocol depicted in \Cref{fig:protocol}}
  \label{tab:roles}
  \smaller
  \begin{tabular}{L{.165\linewidth}@{~~}r@{~~}L{.355\linewidth}@{~~}L{.305\linewidth}}
	\toprule
	\bfseries Role
		& $\boldsymbol{\#}$~
		& \bfseries Tasks
		& \bfseries\itshape Remarks
	\\
	\midrule
	\textbf{Assessors}
		& $\geqslant 2$
		& Peruse the sources to choose one or two categories
		  for each library
		& \itshape An assessor's work must be read only by arbitrators
	\\
	\midrule
	\textbf{Arbitrator}
		& $\geqslant 1$
		& From the categories chosen by the assessors, select one final category
		  for each library
		& \itshape Role incompatible with ``\emph{Assessor}'';
		  a library is arbitrated by up to one person.
	\\
	\bottomrule
  \end{tabular}
\end{table}

While the complete protocol is distributed as several files in the accompanying dataset (see \Cref{sec:results:artifact}), all information needed to understand it is provided in this work:
\Cref{tab:roles} describes the roles of the actors involved (the second column indicates the number of actors required per role);
\ref{app:guidelines} shows a summarised version of the main guidelines that the actors must follow;
\ref{app:categories} shows the natural-language descriptions that characterise the \PyPI Topics used as taxonomical functional categories.

\smallskip
Thus, following our protocol, three or more actors can categorise libraries from any ecosystem with \PyPI Topics, by performing the steps from \Cref{fig:protocol} that we detail next and illustrate with concrete examples from our demonstration.

\setcounter{subsection}{-1}
\subsection{Libraries definition}
\label{sec:protocol:libraries}

Initially, the libraries to be categorised must be defined according to some unambiguous criteria.
We study security aspects of vulnerable \FOSS libraries: for this work we focus on Java libraries available in Maven Central. 
Besides the ecosystem, we selected libraries by: 
\begin{itemize}[leftmargin=1.5em,parsep=0pt,topsep=.7ex,itemsep=.5ex]
\item
\emph{severity of known vulnerabilities},
for which we queried the database of the Snyk website \cite{SnykDB}, looking for \gav in Maven with one or more \CVEs whose $\CVSS\geqslant7$.
\item
\emph{source code availability},
choosing only those libraries for which the \gav could be downloaded from the repositories listed in the corresponding Maven entry.
\end{itemize}

In particular, we used the interface of Snyk to perform this initial security filter%
\footnote{%
The \CVSS values from Snyk and those taken from the National Vulnerability Database (\NVD) occasionally differ. 
Since the \NVD is the authoritative source of \CVE data, the final dataset exhibits the \CVSS from the \NVD.
This resulted in 13\% of the cases---i.e.\ 65 out of 434 \CVEs---with a $\CVSS<7$.
}\!.
Combined with the Maven source-code availability criterion, this resulted in 256 libraries which are the outcome of step~\sref{sec:protocol:libraries} in our demonstration of the protocol, and that we provide in the accompanying dataset (in the \code{data/} directory, see \Cref{sec:results:artifact}).

\subsection{Sources definition and collection}
\label{sec:protocol:sources}

The sources of information, used to interpret the functionality of a library for its categorisation, must be standardised to reduce the subjectivity of the results.
In other words, \emph{the data used to choose a category for a library must be the same for all assessors, and as consistent as possible across the different libraries.}
Therefore, after defining the corpus of libraries, the next protocol stage is to define and collect all data sources related to those libraries that will be used in the subsequent categorisation.

\smallskip
The nature of the sources depends on the ecosystem---they must provide information of the functional purpose of the libraries under study.
This is based on the concept of \emph{data triangulation} \cite[p.~116]{Yin09}, which must be performed until reaching the \emph{saturation} point recommended for qualitative analysis \cite{CS15}.
For our Java/Maven demonstration this resulted in four sources per library:
\begin{enumerate}[label={(\parbox[b]{.6em}{\centering\alph*})},
                  leftmargin=2em,parsep=0pt,itemsep=.3ex,topsep=.5ex]
\item	the \ga main entry in Maven Central,
		which typically contains a brief description of function and purpose;
\item	the main repository of the \ga, e.g.\ in GitHub, GitLab,
		or another git or subversion repository;
\item	the main website of the \ga
		---this can also be an entry for the artifact (\tta)
		within the website of the group (\ttg);
\item	a wiki or documentation page that describes the use of the \ga,
		which could be e.g.\ a dedicated wiki, a Wikipedia entry,
		or a section in the website documenting the \ga in detail---links from
		the same domain than the \ga website (or GitHub) are preferred over
		external sources like Wikipedia.
\end{enumerate}

We originally started with the first two sources only, but data triangulation led us to increase it to the four listed above.
In particular, some sources do not exist for certain libraries, like a main repository, while others may contain scarce information, e.g.\ some Maven Central snippets just repeat the name of the \ga.

We did not continue adding sources because the four above provided multiple information points of the functional purpose for every library in our dataset---that is, triangulation reached saturation.
Adding more sources, such as an additional documentation page or Wikipedia entry, would result in no gain to data quality.

\smallskip
The results of this step must be recorded in a document to which all actors have access, and where at most one web link is provided per source per library.
When a source is not found for a library, the corresponding link space is left blank. 
In our demonstration we found four sources for most libraries, but in a few cases only three were found, because no clear references could be traced to: the main repository of the \ga (6/256 cases), its website (13/256), or any wiki/doc page (18/256).

\subsection{Libraries assessment (categorisation)}
\label{sec:protocol:assess}

Once the sources for a library have been collected, each assessor must revise them to choose a category which best reflects its main functional purpose.
All categories and our minimal descriptions for them are in \ref{app:categories}.
Additional rules of thumb to help in the assessment are in the files \texttt{protocol.ods} and \texttt{.txt} in our dataset \cite{PFMB24}.

\subsubsection{Categorisation by human subjects}
\label{sec:protocol:assess:category}

This is one of the most sensitive steps of the protocol, as it depends on the inference of an abstract function from human-written descriptions.
While human actors still surpass any existent technology at this task, e.g.\ \ML or \NLP algorithms, the results are inherently subjective.
Therefore, various measures were implemented to control such subjectivity, and different justified interpretations are disambiguated in the last protocol step. 

First of all, categories are provided together with subcategories, the latter also taken from \PyPI.
This helps to identify certain functions pertaining to a category.
For instance, category \texttt{Office/Business} has the subcategories
\emph{Financial}, 
\emph{Groupware}, 
\emph{News/Diary}, 
\emph{Scheduling}, and
\emph{Office Suites};
instead, \code{Sociology} has
\emph{Genealogy} and
\emph{History}.

Assessors must consult preselected sources only.
For instance, for \texttt{cd.go.plugin:go-plugin-api} the wiki/doc link is its official API documentation\footnote{\url{https://api.gocd.org/21.1.0/\#introduction}}.
As a consequence, assessors will not consult Wikipedia or any other documentation to decide the category for that library (although they will consult the other three preselected source links).
Notwithstanding, assessors are allowed to investigate technical terms unfamiliar to them that appear in such sources.
For example, if a \ga is described in a source as an ``\acronym{hdds} implementation'', an assessor who does not know the acronym can consult other material to find out that it means ``Hadoop distributed data store''.

Moreover, brief textual descriptions were produced for each category to guide assessors in their inference task.
For example, the \code{Text~Editors} category should contain ``\textit{Libraries used for basic text editing, such as Emacs, Vim, notepad, sublime, etc. Office-specific word processors (like Microsoft Word) don't fall here, but go to \texttt{Office/Business} instead}''.
Such descriptions for all categories are provided in our dataset, and here in \ref{app:categories}.

Finally, some rules-of-thumb with sample scenarios were generated.
They simulate some of the choices that an assessor may face, and what is the expected reaction.
For instance, it is indicated that ``\textit{\texttt{System} and \texttt{Utility} are quite overarching categories that can be confused between each other}''.
The following is an excerpt of the rules of thumb for this case:

\medskip
\noindent%
\fcolorbox{black}{black!4}{%
	\slshape%
	\smaller%
	\begin{minipage}{.96\linewidth}
	\begin{enumerate}[leftmargin=2em,label=(\alph*),parsep=0pt,topsep=.5ex]
	\item
	\label{rule:sys1}
	If the purpose of the library can be identified with services from an Operating System (e.g.\ file manipulation, inter-process communication, binary execution like in runtime environments, process scheduling, firmware or hardware control/interface), then \texttt{System} is the best match.
	\end{enumerate}
	[\,\ldots] ~Example~\ref{rule:sys1}: \\
	%
	The library \texttt{com.github.fracpete:vfsjfilechooser2} says in the sources that it ``is a mavenized fork of the dormant vfsjfilechooser project'', which in turn is ``a File browser for Apache Commons VFS version 2''.
	All these functions are within the scope of file manipulation, so this library can be categorised as \texttt{System}.
	\end{minipage}
}

\medskip
The sources describe the intended purpose of the library from multiple perspectives:
After reading them, an assessor chooses the best matching category in accordance to the protocol guidelines.
However, it is possible that two categories seem equally applicable.
Language markup libraries offer an example: if they are mainly used to process web content, then according to the protocol guidelines they could be categorised as \code{Text} \code{Processing} but also as \code{Internet}.
In those cases the assessor can register both categories, and the final arbitration step will decide which one to choose, comparing them also to the choice(s) made by other assessors---see \Cref{sec:protocol:arbitrate}.

Nevertheless, while category multiplicity improves the versatility of the resulting dataset, imposing a hard limit is important: \emph{we allow up to two choices per library per assessor to keep the intended semantics of using PyPI Topics as categories}.
In contrast, limitless categories would turn into tags, losing the quality of main functional purpose to become instead functional coverage (e.g.\ ``this library also handles audio files, so let's add \texttt{Multimedia}\ldots'').

\subsubsection{Use case: classification of libraries by their category}
\label{sec:protocol:assess:class}

%
\begin{table}
  \centering
  \caption{\parbox[t]{.85\linewidth}{%
	Partition of categorised libraries into classes according to their
	Internet orientation---see \Cref{sec:protocol:assess:class}}}
  \label{tab:classes}
  \smaller
  \begin{tabular}{L{.135\linewidth}@{~~}L{.325\linewidth}@{~}L{.46\linewidth}}
	\toprule
	\bfseries Class
		& \bfseries \!\!\!\mbox{Categories covered}
		& \bfseries Example
	\\
	\midrule
	\bfseries\texttt{Remote network}
		& \parbox[b]{\linewidth}{%
			\code{System},\\
			\code{Database},\\
			\code{Communications},\\
			\code{Security},\\
			\code{Internet},\\
			\code{Utilities}.
		  }
		& Library \code{org.opencastproject:} \code{opencast\text{-} common\text{-} jpa\text{-}impl}, which is a ``database connection interface'', was assigned category \code{Database} and thus falls in class \code{Remote} \code{network}. 
	\\
	\midrule[.4pt]
	\bfseries\texttt{Local}
		& \texttt{Multimedia},
		  \texttt{Text Editors},
		  \texttt{Text Processing},
		  \texttt{Scientific/ Engineering},
		  \texttt{Software Development},
		  etc.
		& Library \code{org.apache.pdfbox:} \code{pdfbox}, which is ``an open source Java tool for working with PDF documents'', was assigned category \code{Text} \code{Processing} and therefore belongs to the class \code{Local}.
	\\
	\bottomrule
  \end{tabular}
  \vspace{.5ex}
  \justifying  

  \noindent
  \textbf{Note:} class \code{Local} includes the 18 \PyPI Topics not listed here for \code{Remote} \code{network}. The list given above for the \code{Local} class is not exhaustive, but includes all categories in our dataset---see \Cref{fig:final_categories}.
  \\[1ex]
  \textbf{Exceptions:} using categories as proxies for the classes  sometimes results in the misclassification of a library.
	\Cref{sec:protocol:assess:class,sec:protocol:arbitrate:class} explain the mechanisms to identify and correct such situations.
\end{table}

Our protocol serves to assign functional categories to libraries.
This creates a fine-grained division, but one may also want to generate more coarse-grained \emph{classes}.
This can be done with no modification to the assessment step, namely by assigning each category to a class.

For example, consider a separation of the libraries into classes \code{Remote~network} and \code{Local}, depending on their relation to computer networks, namely on whether they are typically used for activities with high exposure to the Internet.
Then, researchers can agree on a partition of the categories as the one shown in \Cref{tab:classes}.

However, a classification proxied by categories can sometimes be imprecise: while it may handle well most cases, it is possible for the final category of a library to be misaligned with its class.
This occurs when classes are defined on concepts that are orthogonal to the functional division of some categories.
In \Cref{tab:classes} this can happen e.g.\ for \code{Text} \code{Processing} libraries, that \emph{in general} belong to the class \code{Local}, but may fit better the class \code{Remote} \code{network} when applied to markup parsers dedicated to server responses.

The template engine \code{com.hubspot.jinjava:jinjava} offers a concrete example of the above.
This library was categorised in our demonstration as \code{Software~Development}, which belongs to the \code{Local} class in \Cref{tab:classes}. 
However, \code{jinjava} is a template specialised on network-related software, so \code{Remote} \code{network} could be the best class.

Such situations call for a systematic solution, that ensures a coherent decision process for the different libraries.
For instance, the protocol could be refined with a class-revision phase during the final arbitration step, in which ``suspicious'' libraries would be examined under the scope of the class-division rule (Internet orientation in our example).
This solution is practical as long as class-revision needs to be performed on a subset of the libraries only.
As it involves the arbitration step, we discuss it next in \Cref{sec:protocol:arbitrate}, including a practical implementation example.

\subsection{Final arbitration}
\label{sec:protocol:arbitrate}

Once all assessors have chosen the categories for a library $\ell$, if these assessments are ambiguous, then the arbitrators must define a single final category for $\ell$---because our purpose is to determine a proper case-control study across ecosystems \cite{AM14,VWM16}.
Alternatively, when the assessors agree on the categories of $\ell$, no arbitration is needed.

\subsubsection{Non-binary ambiguity}
\label{sec:protocol:arbitrate:category}

The above requires ambiguity to be a Boolean property.
But that is not necessarily true in the general case, as our demonstration exemplifies, because partial agreements can result from $K>1$ assessors that choose $N\geqslant1$ categories for a library $\ell$.
A point-wise metric is thus required, to quantify the ambiguity of each individual library assessment---this in contrast to traditional statistical measures of agreement, such as Cohen's and Fleiss' kappa, that measure global agreement among assessors (and that we use to evaluate the complete dataset in the next \namecref{sec:results}).

The protocol solves the requirement of such point-wise metric by quantifying the ambiguity of every assessment via its \emph{conflict} value, defined for each library $\ell$ as the ratio of different assessments for $\ell$ to the number of explicit choices made for it by all assessors:
\begin{align}
	\conflict(\ASS_\ell)
		&= \frac{\misses(\ASS_\ell)}{\choices(\ASS_\ell)}
		\nonumber
		\\[.5ex]
		&= \frac{\choices(\ASS_\ell) - K\cdot\matches(\ASS_\ell)}%
		        {\choices(\ASS_\ell)}
		\nonumber
		\\[.5ex]
		&= \frac{\sum_{i=0}^{K-1}{\card{\ASS_\ell[i]}}
		        	- K\card{\bigcap_{i=0}^{K-1}\ASS_\ell[i]}}%
		        {\sum_{i=0}^{K-1}{\card{\ASS_\ell[i]}}}
	\label{eq:conflict}
\end{align}
where $\card{A}$ is the cardinality of set $A$, and $\ASS_\ell$ is an array of length $K$ whose $i$-th position $\ASS_\ell[i]$ is the set of categories chosen by the $i$-th assessor for library $\ell$.

The codomains of functions $\choices(\cdot)$ and $\matches(\cdot)$ in \Cref{eq:conflict} are respectively $\{1,2,\ldots,KN\}$ and $\{0,1,\ldots,N\}$.
Therefore, for arbitrary $K$ and $N$ the function $\conflict(\cdot)$ can take any value in the rational interval $[0,1]$.
This allows for a Boolean definition of ambiguity, by choosing a threshold $T\in[0,1)$ and calling library $\ell$ ambiguous if $\conflict(\ASS_\ell)>T$.

In our setting with $K=2$ assessors and up to $N=2$ categories per library, $\conflict(\cdot)$ can take four different values---see \Cref{tab:conflict}.
We identify the absence of coincidences among all assessments via the threshold $T=\sfrac{1}{2}$, obtaining the Boolean concept of ambiguity illustrated in the rightmost column of the \namecref{tab:conflict}.
Together with our requirement to have a single final category per library, this results in three possible arbitration scenarios:
\begin{enumerate}[leftmargin=2em,parsep=0pt,topsep=1ex,itemsep=1ex]
\def\mybullet{\raisebox{1pt}{\ensuremath{\scriptstyle\blacktriangle}}}
\item
\emph{The case is not ambiguous, and there is a single coincident category among the choices of all assessors}.
Then no arbitration is needed and that category is selected as final for the library.
	\begin{itemize}[label=\mybullet,leftmargin=1.3em,parsep=0pt,topsep=.7ex]
	\item  In our Java/Maven demonstration this happened to 148 libraries, which amounts to 57.8\% of the total---see \Cref{fig:piecharts:ambiguity} in \Cref{sec:results}.
	\end{itemize}
\item
\emph{The case is not ambiguous, and there are two coincident categories}.
Here, arbitration must choose between those two categories, i.e.\ no third category is possible.
Other than that, the tasks to perform are equivalent to those of the assessors: the arbitrator must read the sources and select \emph{one} final category for the library, according to the predefined rules.
	\begin{itemize}[label=\mybullet,leftmargin=1.3em,parsep=0pt,topsep=.7ex]
	\item	This happened 4 times in our dataset.
	\end{itemize}
\item
\emph{The case is ambiguous (no coincident categories)}.
Arbitration in this case is equivalent to that in the previous point: the set of categories to choose from is given by the union of categories chosen by all assessors.
	\begin{itemize}[label=\mybullet,leftmargin=1.3em,parsep=0pt,topsep=.7ex]
	\item	This happened to 104 libraries in our dataset, i.e.\ 40.6\% of the total---see \Cref{fig:piecharts}.
	\end{itemize}
\end{enumerate}

%
\begin{table}
  \def\ln#1{\texttt{\smaller[.5]\color{Gray}#1}}
  \centering
  \caption{Quantification of conflicts in a library assessment}
  \label{tab:conflict}
  \smaller[.5]
  \vspace{-.5ex}
  \begin{tabular}{L{9pt}@{~}cc cc cc}
	\cmidrule[.8pt](l{-3pt}){2-7}
	&	  \multicolumn{2}{@{~}c}{Ass.~$A$}
		& \multicolumn{2}{c}{Ass.~$B$}
		& 
	\\
	\cmidrule(l{1pt}r){2-3}
	\cmidrule(lr){4-5}
	&	  $A_1$ & $A_2$
		& $B_1$ & $B_2$
		& \multirow{-2}{*}{$\conflict(\tuple{A,B})$}
		& \multirow{-2}{*}{Ambiguous?}
	\\
	\cmidrule[.5pt](l{-3pt}){2-7}
	\ln{1} & $x$   & $ $   & $x$   & $ $    & $0$            & $\bot$ \\
	\ln{2} & $x$   & $y$   & $x$   & $y$    & $0$            & $\bot$ \\
	\ln{3} & $x$   & $y$   & $x$   & $ $    & $\sfrac{1}{3}$ & $\bot$ \\
	\ln{4} & $x$   & $y$   & $x$   & $z$    & $\sfrac{1}{2}$ & $\bot$ \\
	\ln{5} & $x$   & $ $   & $y$   & $ $    & $1$            & $\top$ \\
	\ln{6} & $x$   & $y$   & $z$   & $ $    & $1$            & $\top$ \\
	\ln{7} & $x$   & $y$   & $z$   & $w$    & $1$            & $\top$ \\
	\cmidrule[.8pt](l{-3pt}){2-7}
  \end{tabular}
  \vspace{1ex}

  \noindent
  \explanation{%
Assessor $A$ chooses categories $A_1$ and $A_2$ for the library, idem for $B$.
Blanks stand for no choice: e.g.\ on line \ln{3}, $A$ chooses categories $\{x,y\}$ while $B$ chooses only $\{x\}$.
The as\-sessments' conflict value is given by $\conflict(\cdot)$ from \Cref{eq:conflict}.
An assessment is ambiguous if ${\conflict(\tuple{A,B})>\sfrac{1}{2}}$. 
  }
\end{table}

\subsubsection{Use case: revision of library classes by arbitrators}
\label{sec:protocol:arbitrate:class}

We take up on the use case from \Cref{sec:protocol:assess:class}, where the coarser-grained concept of class was defined over categories.
This introduced the need for an extra phase in the arbitration step, where arbitrators must revise the final class to which a library belongs, e.g.\ \code{Local} vs.\ \code{Remote} \code{network}.
However, the exercise is practical only if the workload generated is lower than a reassessment of the libraries under the class-division rule.


One practical way to systematise such process is to leverage data available from previous steps, that has information relevant to the class-division rule, and allows for an automatic marking of the cases that will be in need of class revision.

For a concrete example consider our demonstration for step~\sref{sec:protocol:libraries} of the protocol, 
where we selected libraries with high or critical vulnerabilities, i.e.\ with a \CVE whose \CVSS is above 7.
Our dataset thus contains \CVSS data for every library, and the Attack Vector metric of a \CVSS entry ``\textsl{reflects the context by which vulnerability exploitation is possible}''\footnote{\url{https://www.first.org/cvss/v3.1/specification-document}}.
Therefore we can leverage this data, because we are interested in Internet-orientation, and if the Attack Vector is equal to {\guillemotleft}\textsl{Network}{\guillemotright} then ``\textsl{The vulnerable component is bound to the network stack [\,\ldots]  including the entire Internet}'' \cite{MSR06}.

We dub the above \texttt{AV:N}, and recall that a library should belong to class \code{Local} if it is not expected to have high exposure to the Internet.
Therefore, a library that was (automatically) classified as \code{Local} but also has \texttt{AV:N} is worth a second look.
In practice in our demonstration, \code{Local} libraries that have \texttt{AV:N} are marked for class-revision, which amounts to less than 20\% of the dataset.
Arbitrators must then revise these libraries under the class-division concept, which involves reading their sources and \CVE descriptions.
The class can finally be changed to \code{Remote} \code{network} if it is found that the library is in fact used in activities with high exposure to the Internet.


Library \texttt{edu.stanford.nlp:stanford-corenlp} offers a concrete example of this procedure.
Its sources indicate that it was developed in the University of Stanford to provide ``\textit{a set of natural language analysis tools written in Java}''.
This \ga was assigned the final category \code{Scientific/Engineering} in our demonstration, which belongs to class \code{Local}.
However, since the \CVEs related to this library have \texttt{AV:N}, it was marked for the class-revision process.
It was then found that the \CVEs ``\textit{describe XML vulnerabilities of the server which this artifact DOES implement}'', which convinced the arbitrator to change the class of this library to \code{Remote} \code{network} \cite{PFMB24}.

\section{Resulting dataset}
\label{sec:results}

\begin{figure}
	\centering
	\includegraphics[width=.95\linewidth]{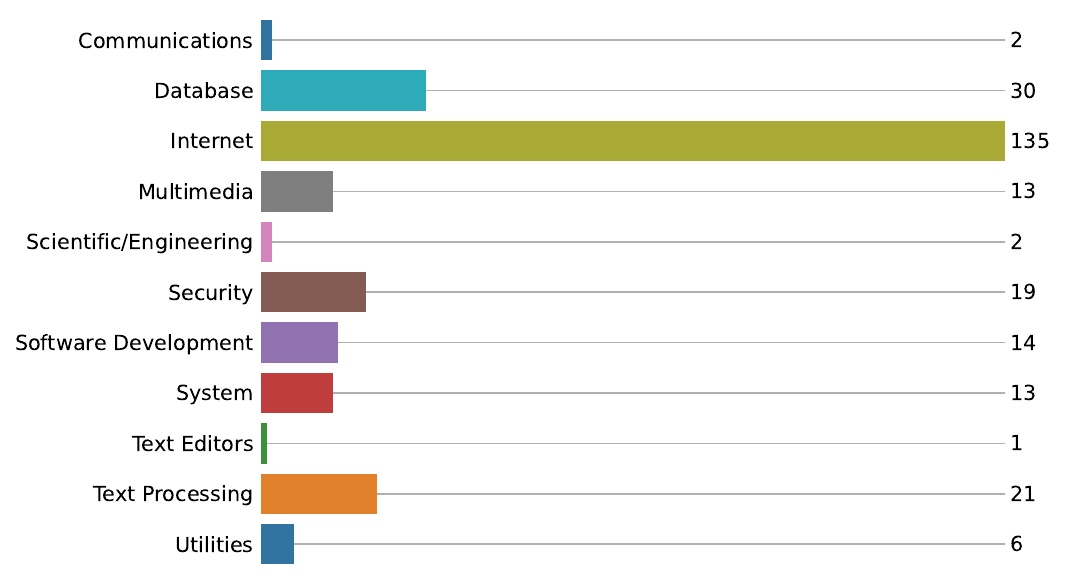}
	\explanation{\smallskip%
Categories in our dataset are not balanced---this was expected, given our selection of Java libraries that had security vulnerabilities with high- or critical-severity \CVSS.
The 13 \PyPI categories missing from this chart had no libraries assigned in our dataset.%
}
	\caption{Number of Maven libraries assigned to each \PyPI Topic}
	\label{fig:final_categories}
\end{figure}

We implemented the protocol from \Cref{sec:protocol}, demonstrating its use to categorise libraries from the Java/Maven ecosystem with \PyPI Topic classifiers.
We used three people, namely two assessors and an arbitrator, pipelining work via online spreadsheets (e.g.\ a {\guillemotleft}\textsl{Done}{\guillemotright} mark that assessors would set for a library when the corresponding categories had been chosen).
To avoid having fast-paced assessors influence slower ones, or any other inter-actor bias, it suffices to enforce the read permissions indicated in \Cref{tab:roles}.
With our minimal team, this pipelining helped to speed up the categorisation of the 256 libraries.

\subsection{Dataset summary}

\begin{figure}
	\centering
	\begin{subfigure}{.47\linewidth}
		\centering
		\raisebox{3pt}{%
		\includegraphics[width=.85\linewidth]{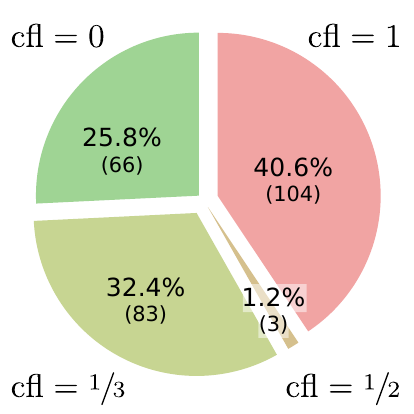}}
		\caption{\parbox[t]{.85\linewidth}{\RaggedRight%
			Quantification of assessment conflicts via \Cref{eq:conflict}}}
		\label{fig:piecharts:conflict}
	\end{subfigure}
	~
	\begin{subfigure}{.47\linewidth}
		\centering
		\includegraphics[width=\linewidth]{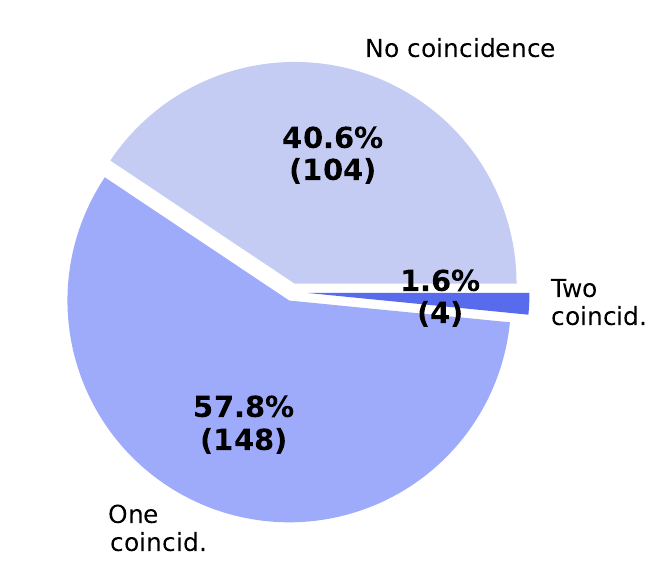}
		\caption{\parbox[t]{.85\linewidth}{\RaggedRight%
			Arbitration needed on $40.6\%+1.6\%$ of the cases}}
		\label{fig:piecharts:ambiguity}
	\end{subfigure}
	\explanation{\bigskip%
Assessors agreed in the majority of the cases on one suggested category (57.8\%), and in four cases even on both possible categories (1.6\%).
Inter-rater reliability Fleiss' $\kappa=0.382$ 
suggests fair agreement among assessors.
}
	\caption{Agreement among assessors' categorisation}
	\label{fig:piecharts}
\end{figure}

The final results are presented in \Cref{fig:final_categories,fig:piecharts,tab:results}.
We highlight that 53\% of the libraries were assigned the category \code{Internet}: this is not surprising, considering our initial filter on libraries whose \CVEs had high- or critical-severity \CVSS.
 In fact, from the 256 libraries, only 10 had no \CVEs with a {\guillemotleft}\textsl{Network}{\guillemotright} Attack Vector metric---we discuss this further in what follows. 

\paragraph{Inter-rater statistics}
This problem is a categorical labelling of data with multiple raters and multiple categories.
Therefore, we use Fleiss' kappa as a measure of inter-rater reliability between the two assessors \cite{FLP03}.
For the technical implementation, in the cases were an assessor chose two categories, we kept only one but respecting any coincidence with the other assessor.
That is, when assessor $A$ chose categories $x$ and $y$ for a library, there are three options regarding the choices of assessor $B$:
(\textit{i})\:if $B$ chose $x$ then we keep $x$ as the category for both assessors;
(\textit{ii})\:if $B$ chose $y$ then we keep $y$ as the category for both assessors;
(\textit{iii})\:if $B$ chose $z$ then we keep $x$ as the category for $A$ and $z$ for $B$.
As a result, we get $\kappa=0.381599$ which indicates fair agreement between the assessors \cite{LK77,FLP03}.

%
\begin{table}
  \centering
  \begin{threeparttable}  
  \caption{Statistics of the categorisation: second column is the number of libraries in the category/class; next columns are statistics on the \CVSS for the \CVEs of those libraries\,\tnote{$\ast$}}
  \label{tab:results}
  \smaller
  \def\sep{\hspace{5.7pt}}
  \setlength{\aboverulesep}{.3pt}
  \setlength{\belowrulesep}{.3pt}
  \setlength{\extrarowheight}{.4ex}
  \begin{tabular}{L{.38\linewidth}@{~}c@{\quad}c@{\sep}c@{\sep}c@{\sep}c@{\sep}c}
	\toprule
		& 
		& \multicolumn{5}{c}{\bfseries \CVSS}
	\\
	\cmidrule{3-7}
	\multirow{-2}{*}{\bfseries Category\tnote{\dag}}
		& \multirow{-2}{*}{~$\boldsymbol{\#}$~}
		& Min
		& Med
		& Max
		& Avg
		& Stdev
	\\
	\midrule
	\arrayrulecolor{black!65}
	\rowcolor{shade1}
	\code{Internet}
		& 135
		& $4.70$
		& $7.50$
		& $10.00$
		& $8.04$
		& $1.29$
	\\
	\midrule[.3pt]
	\rowcolor{shade1}
	\code{Database}
		& 30
		& $5.90$
		& $9.10$
		& $9.80$
		& $8.66$
		& $1.15$
	\\
	\midrule[.3pt]
	\code{Text~Processing}
		& 21
		& $5.30$
		& $8.50$
		& $9.80$
		& $8.03$
		& $1.49$
	\\
	\midrule[.3pt]
	\rowcolor{shade1}
	\code{Security}
		& 19
		& $5.30$
		& $8.45$
		& $9.90$
		& $8.19$
		& $1.39$
	\\
	\midrule[.3pt]
	\code{Software~Development}
		& 14
		& $5.50$
		& $7.50$
		& $9.80$
		& $8.11$
		& $1.32$
	\\
	\midrule[.3pt]
	\code{Multimedia}
		& 13
		& $6.50$
		& $6.50$
		& $10.00$
		& $7.33$
		& $1.14$
	\\
	\midrule[.3pt]
	\rowcolor{shade1}
	\code{System}
		& 13
		& $5.90$
		& $9.00$
		& $10.00$
		& $8.70$
		& $1.25$
	\\
	\midrule[.3pt]
	\rowcolor{shade1}
	\code{Utilities}
		& 6
		& $7.50$
		& $8.10$
		& $9.80$
		& $8.40$
		& $0.63$
	\\
	\midrule[.3pt]
	\code{Communications}
		& 2
		& $6.80$
		& $7.95$
		& $9.10$
		& $7.95$
		& $1.15$
	\\
	\midrule[.3pt]
	\code{Scientific/Engineering}
		& 2
		& $5.30$
		& $7.50$
		& $9.80$
		& $7.53$
		& $1.84$
	\\
	\midrule[.3pt]
	\code{Text~Editors}
		& 1
		& $9.80$
		& $9.80$
		& $9.80$
		& $9.80$
		& $0.00$
	\\
	\arrayrulecolor{black}
	\midrule[1pt]%
	\bfseries Class\tnote{\ddag} \phantom{\textlarger[2]{A}} & & & & & & 
	\\
	\midrule
	\arrayrulecolor{black!65}
	\rowcolor{shade1}
	\code{Remote~network}
		& 211  
		& $4.70$
		& $8.10$
		& $10.00$
		& $8.22$
		& $1.24$
	\\
	\midrule[.3pt]
	\code{Local}
		& 48  
		& $5.30$
		& $7.50$
		& $10.00$
		& $7.92$
		& $1.46$
	\\
	\arrayrulecolor{black}
	\bottomrule
  \end{tabular}
  \footnotesize
  \begin{tablenotes}
	\item[$\ast$] \CVSS values according to the \NVD \cite{NVDDB}, of the pre-selected \CVEs associated to our 256 libraries---see \Cref{sec:protocol:libraries}.
	\item[\dag] The 13 categories missing from this table---e.g.\ \code{Printing} and \code{Terminals}---had zero libraries assigned to them.
	\item[\ddag] Statistics per class are orthogonal to those per category: here, class \code{Remote\ network} includes categories \code{Internet}, \code{Database}, \code{Security}, \code{System}, and \code{Utilities}; the rest are in class \code{Local}. Mismatches in the second column (between classes and categories counts) are due to class revision---see \Cref{sec:protocol:arbitrate:class}.
  \end{tablenotes}
  \end{threeparttable}
\end{table}

\smallskip
Class-revision changed the class of 6 libraries, i.e.\ 2.3\% out of the 256 libraries in our dataset.
\Cref{sec:protocol:arbitrate:class} described how the class of \texttt{edu.stanford.nlp:}\texttt{stanford-corenlp} was changed from \code{Local} to \code{Remote} \code{network} in our demonstration.
In contrast, here we show how class revision \emph{did not change the class} of library \texttt{com.itextpdf:io}.
That \ga was an ambiguous case, in which one assessor chose category \code{Text} \code{Editors} and the other chose \code{Text} \code{Processing}.
The arbitrator resolved the conflict to \code{Text} \code{Editors} and wrote the following comment (sic):

\smallskip
\noindent
\fcolorbox{black}{black!4}{%
	\slshape%
	\begin{minipage}{.96\linewidth}
Library ``io'' provides the ``low-level functionality'' of iText 7 which is a PDF reader and editor;\\ CVE desrcibes command injection via another tool, which is irrelevant for the categorsation of this library
	\end{minipage}
}

\medskip
Since category \code{Text} \code{Editors} belongs to class \code{Local}, and \texttt{CVE-2021-43113} associated to this library has \code{AV:N}, it was marked for class revision.
The arbitrator then considered whether to change the class of this library to \code{Remote} \code{network}, by reading its sources and the corresponding \CVE entry.
The final decision was to keep the class as \code{Local}, which is explicitly indicated in the comment above, after the semicolon.

%
\begin{table*}
  \centering
  \begin{threeparttable}  
  \caption{Columns in the sheet \texttt{libs\_arbitration} of file
  	\texttt{arbitration.ods}, where there is one library per row}
  \label{tab:arbitration}
  \smaller
  \colortablepreamble
  \begin{tabular}{c@{~}L{.57\linewidth}@{~~}L{.36\linewidth}}
	\toprule
	Cols.
	& Description
	& Example from \texttt{libs\_arbitration}
	  (row 65)
	\\\midrule
	A
	& Library \ga coordinates from Maven Central
	& \texttt{edu.stanford.nlp:stanford-corenlp} 
	\\
	B
	& Category that assessor 1 chose for the library
	& \code{Scientific/Engineering}
	\\
	C
	& Alternative category chosen by assessor 1, if any
	& 
	\\
	D
	& Category that assessor 2 chose for the library
	& \code{Internet}
	\\
	E
	& Alternative category chosen by assessor 2, if any
	& \code{Text~Processing}
	\\
	F--G
	& Whether assessor 1,2 is done (for the pipelined process,
	  see \Cref{sec:protocol:arbitrate})
	& DONE
	\\
	H--K
	& Formulae to compute the conflict metric following \Cref{eq:conflict}:
	  $\conflict(\langle \{\text{B,C}\},\{\text{D,E}\} \rangle)$=
	& 1
	\\
	L
	& Ambiguous assessment? ``YES'' if $\conflict(\cdots)>T=\sfrac{1}{2}$,
	  ``NO'' otherwise
	& YES
	\\
	M
	& Category on which both assessors coincide
	  (if col.\:L is ``NO''; blank otherwise)
	&
	\\
	N
	& Category chosen by the arbitrator
	  (if col.\:L is ``YES'', blank otherwise)
	& \code{Scientific/Engineering}
	\\
	O
	& Final category for the library, i.e.\ disjoint union of
	  cols.\:M and N
	& \code{Scientific/Engineering}
	\\
	P
	& Class to which the category from col.\ K belongs,
	  i.e.\ \code{Local} or \code{Remote~network}
	& \code{Local}
	\\
	Q
	& ``YES'' when a \CVE of the library has \texttt{AV:N}, ``NO'' otherwise
	  (for class revision)\tnote{$\ast$}
	& YES
	\\
	R--S
	& Revised class: it can equal the class of the category, or be
	  \code{Remote~network}
	& \code{Remote~network}
	\\
	T
	& Natural-language comment on the decisions by the assessors
	  and arbitrator
	& \textit{Artifact ``Provides a set of natural language [\,\ldots]%
		; \CVEs describe XML vulnerabilities of the server which this artifact
		DOES implement [\,\ldots]
		}
	\\\bottomrule
  \end{tabular}
  \footnotesize
  \begin{tablenotes}
	\item[$\ast$]
	Formula over sheet \code{libs\_CVEs\_AV}, that checks whether any \CVE
	column for the library reads \code{NETWORK}---see \Cref{sec:results:artifact}.
  \end{tablenotes}
  \end{threeparttable}
\end{table*}

\subsection{Open-data artefacts}
\label{sec:results:artifact}

\noindent
We provide files for every artefact produced in this work: 

\begin{fileslist}
\item	\label{files:protocol}
\emph{the protocol}
overview (\acronym{svg} and \acronym{pdf} images), and detailed description (spreadsheets and ASCII text files) with natural-language information about:
\begin{itemize}[label=\raisebox{1.3pt}{\smaller[3]$\blacktriangleright$},
                leftmargin=1.5em,parsep=0pt,topsep=.5ex]
\item	\emph{definitions of roles}, number of people per role,
		compatibility across different roles for one actor,
\item	\emph{the set of categories}, plus subcategories
		and the partition into classes \code{Remote~network}
		and \code{Local},
\item	\emph{steps of the protocol} as per \Cref{fig:protocol},
		with substeps such category arbitration and class revision,
\item	\emph{guidelines for assessors} to help in the process of
		choosing the category for a library, including the
		rules-of-thumb mentioned in \Cref{sec:protocol:assess};
\end{itemize}
\item	\label{files:libs}
\emph{the list of 256 libraries},
as an ASCII text file with \ga coordinates (one per line)
from Maven Central;
\item	\label{files:sources}
\emph{the sources} collected for each library:
\begin{itemize}[label=\raisebox{1.3pt}{\smaller[3]$\blacktriangleright$},
	        	leftmargin=1.5em,parsep=0pt,topsep=.5ex]
\item	\emph{web links} as originally used in our work,
		in a spreadsheet with one library and its sources per row
\item	\emph{\acronym{pdf} printouts} of the corresponding web pages;
\end{itemize}
\item	\label{files:assessments}
\emph{the assessment sheets} of both assessors;
\item	\label{files:arbitration}
\emph{the arbitration sheet}, which includes:
\begin{itemize}[label=\raisebox{1.3pt}{\smaller[3]$\blacktriangleright$},
		        leftmargin=1.5em,parsep=0pt,topsep=.5ex]
\item	the final category and revised class per library,
\item	the \CVEs (and their \NVD \CVSS and \acronym{av} metrics)
		considered for this study,
\item	natural-language comments on the reasons for choosing
		the categories, or arbitration where applicable.
\end{itemize}
\end{fileslist}

We further provide files with aggregated information on all data gathered and generated, including the scripts used to query web sources (e.g.\ the \NVD database), and to process the data in order to generate the figures and tables of this article.
These scripts have been prepared to run directly on the directory tree structure where they are provided, so e.g.\ executing \code{stats\_categories.py} reads the data from \code{data/arbitration.csv} and produces:
\begin{itemize}[parsep=0pt,topsep=1ex] 
\item	the categorisation-counts bars charts (\Cref{fig:barschart,fig:final_categories}),
\item	the assessment-\-coincidences pie chart (\Cref{fig:piecharts:ambiguity}), and
\item	the inter-rater statistic (Fleiss' kappa value).
\end{itemize}

All files have been uploaded to an online repository.
They are publicly shared under a \acronym{cc-by-sa} 4.0 license---or GPLv3 in the case of the PyPI scripts---available as \cite{PFMB24} and under the following link%
\footnote{\textbf{\sffamily\color{Red}NOTE to reviewers:}
the dataset is currently shared under restricted access for reviewing purposes only---it will be made public upon acceptance of the manuscript.}
:
\par\nopagebreak\medskip
\begin{centering}
\begin{minipage}{\linewidth}
	\centering
	\smaller
	\fcolorbox{blue}{orange!3}{%
		\href{https://zenodo.org/records/10480832}%
		{\texttt{https://zenodo.org/records/10480832}}
    }\\[-1.3ex]%
	\begin{forest}
	for tree={
	  font=\ttfamily,
	  grow'=0,
	  anchor=west,
	  child anchor=west,
	  parent anchor=south,
	  calign=first,
	  inner xsep=7pt,
	  inner ysep=1.3pt,
	  edge path={
	    \noexpand\path [draw, \forestoption{edge}]
	    (!u.south west) +(7.5pt,0) |- (.child anchor) pic {folder} \forestoption{edge label};
	  },
	  before typesetting nodes={
	    if n=1
	      {insert before={[,phantom]}}
	      {}
	  },
	  fit=band,
	  before computing xy={l=15pt},
	}  
	[
	  [summary.ods, is file]
	  [data]
	  [protocol]
	  [software, inner ysep=.5pt]
	]
	\end{forest}
\end{minipage}
\end{centering}

From the list given above, \namecrefs{files:protocol} for~\ref{files:protocol} are located in the \code{protocol/} directory, while \crefrange{files:libs}{files:arbitration} are inside \code{data/}.
In turn, \code{software/} contains all Python scripts used to process the data, e.g.\ \code{stats\_categories.py} mentioned above, and \code{NVD\_CVSS.py} that queries the \CVSS values (via the \NVD \acronym{api}) used to generate \Cref{tab:results}.

\medskip
Our dataset is further garnished with \code{README} files that explain its directory tree structure and describe all main files included.
The arbitration sheet \code{arbitration.ods} is arguably the most complex and information-rich among them: we briefly describe it next.

The file contains five inter-dependent sheets.
The first one, \code{libs\_arbitration}, shows the main results of the assessments and the arbitration processes, including the class revision.
Each Maven library occupies a row: its \ga coordinates are in column A, while columns B--T condense all information produced for that library---\Cref{tab:arbitration} shows an example for \texttt{edu.stanford.nlp:stanford-corenlp}.

The other four sheets contain data gathered for the libraries, used to compute values used in \code{libs\_arbitration}.
The order of the rows across all sheets is coherent, e.g.\ row~32 corresponds to library \texttt{com.jfinal:jfinal} across all sheets in that file.
The data in the extra sheets is:
\begin{itemize}[leftmargin=1.5em,parsep=0pt,topsep=.5ex]
\item
\code{libs\_CVEs}: \CVEs extracted from the Snyk database during our library-selection process (step~\sref{sec:protocol:libraries} of the protocol).
There is one \CVE per column from B onwards.
	\begin{itemize}[leftmargin=1.5em]
	\item	E.g.\ row~32 for library \texttt{com.jfinal:jfinal} reads \texttt{CVE-2021-31649} in column~B---i.e.\ cell (32,B)---and \texttt{CVE-2019-17352} in cell (32,C).
	\end{itemize}
\item
\code{libs\_CVEs\_AV}: Attack Vector metrics 
of the \CVEs.
	\begin{itemize}[leftmargin=1.5em]
	\item	E.g.\ \code{com.jfinal:jfinal} reads \code{NETWORK} in both cells (32,B) and (32,C), indicating that both \CVEs of that library have \texttt{AV:N}.
	\end{itemize}
\item
\code{libs\_CVEs\_CVSS}: \CVSS values of the \CVEs
\footnote{Statistics for these values are presented in \Cref{tab:results}.}\!.
	\begin{itemize}[leftmargin=1.5em]
	\item	E.g.\ \code{com.jfinal:jfinal} values $9.8$ and $7.8$ in cells (32,B) and (32,C) respectively correspond to CVE-2021-31649 and CVE-2019-17352.
	\end{itemize}
\item
\code{categories\_of\_GAs}: miscellaneous  validation data.
\end{itemize}

\section{Conclusions}
\label{sec:conclu}

We have introduced a cross-ecosystem, systematic, hu\-man-driven protocol for the categorisation of software libraries by functional purpose for SE research.
This work describes the protocol in a manner applicable to any software ecosystem.
We detail the tasks of multiple assessors, who work independently to choose the first-hand library categories, and arbitrators, in charge of aggregating all choices and resolving conflicts.
\emph{This generates the ground truth required as input by most automatic classifiers}, that work on data about the functional purpose of code.

We demonstrated the effectiveness of our protocol, implementing it for the Maven ecosystem by selecting, categorising, and classifying 256 Java libraries.
These were chosen based on the existence of one or more versions having a \CVE whose severity score is High or Critical.
That is, for every \ga in our dataset there is a version \ttv in Maven such that \gav is registered with $\CVSS \geqslant 7$ in the Snyk database \cite{SnykDB}.
This makes our resulting dataset particularly relevant for cybersecurity analyses, e.g.\ studying relationships between library functionality and \CVSS value.


The names of the categories in our protocol come from popular and broadly used functional-purpose classifiers: the (top-level) \PyPI Topics.
This results in a categorical partition of the libraries akin to a functional taxonomy, as opposed to organic and language-specific categorisations that are difficult to extrapolate across ecosystems.
Our use of a functional taxonomy reduces the redundancy and fragmentation of the final set of categories, with benefits to the transparency of choices made by SE researchers, especially when these involve multiple ecosystems.

Besides this fine-grained categorisation, our protocol describes the definition of a higher-level classification that leverages the categories. 
We demonstrate this concept in our implementation, defining the classes \code{Local} and \code{Remote} \code{network} (regarding the position of a library with respect to computer networks).
We applied this classification and performed a class-revision process, that resulted in six libraries having their class changed.

Our procedure, as well as the artefacts generated during our demonstrative implementation, are made freely available in a public repository \cite{PFMB24}.
This is a contribution to the SE community, which can deploy its own implementations for any ecosystem, or directly use the data generated by us for Java/Maven in data-intensive studies.

\subsection{Limitations and Extensions}
\label{sec:conclu:limitations}

Our methodology can be used to build a ground-truth for \ML- or \NLP-based analyses of software properties, such as correlations between the function of a library and other \SE aspects and metrics like popularity, code activity, lines of code, etc.
The actual results of our initial demonstration in \Cref{sec:results} can be the core base to be used for training and validation of such analyses.
Moreover, since the category names used are \PyPI Topic classifiers, cross-language studies can bootstrap from our dataset with no need to repeat the process for Python counterparts.

Our use of a single Topic for each Java library---as unique functional category for it---results in a partition of the libraries that are classified in the target ecosystem.
This allows to determine a proper cross-ecosystem case-control study \cite{AM14,VWM16}.
We note however that some libraries in our implementation had alternative categories assigned by the assessors: we make these available in our dataset---see \Cref{sec:results:artifact}---but highlight that a single category was ultimately chosen per library.
This means that our final dataset can be compared against Maven data:
\begin{itemize}[leftmargin=1.5em,topsep=.5ex,parsep=0pt]
\item
In some cases our categories are less specific than Maven, e.g.\ the Maven category for library \texttt{org.apache.camel: camel-core} is \texttt{Enterprise} \texttt{In\-teg\-ra\-tion}, while we assigned it the \PyPI Topic \texttt{Software} \texttt{Development}.
However, for this reason our categorisation also remains easier to use in cross-\-ecosystem studies.
\item
Our protocol always assigns one category to a library, while Maven has no category assigned to several of the libraries in our dataset.
However, this approach does not create a bijection between Java/Maven and Python/\PyPI libraries, since the latter can potentially have multiple or no Topic classifier assigned.
\end{itemize}

\smallskip
In general, category assessment and arbitration by human subjects is unavoidably subjective, with threats to validity such as construct under-representation, irrelevant variance, and personal bias \cite{Mes89,NS19,MSS+22}.
That is the reason why several guidelines and simulated scenarios are a strong part of the protocol.
The consensed generation and application of these rules is paramount to reduce the final subjectivity of the results.
Our team was small, making it more vulnerable to individual bias---however, our measurements of inter-rater statistics suggest that the categories assessment achieved in the end was not unreliable.

The cost to pay for these results is a long process:
for each library---in our application on Java/Maven---each assessor and arbitrator had to read up to four documents, to then infer the category that describes best the main function(s) of that library.
Therefore, the practical balance between reducing subjectivity (e.g.\ by adding more assessors) and increasing throughput (e.g.\ by partitioning the libraries set and distributing the load among assessors) should be considered before starting the process.

By construction, the size of manual datasets that can be produced in this way grows linearly with the number of people involved in it.
Given the limited size of our team, the generated dataset cannot be \emph{directly} used  for automated large-scale \SE studies.
Instead, \emph{ML and NLP approaches can be used to scale up these results to a much larger number of libraries, for example by using our dataset as a ground truth for partial-label learning \cite{FLH+20}}.


\medskip
As continuation, our protocol could be applied to other ecosystems such as JS/npm or Ruby/Gems, to see whether the high-\CVSS value condition replicates the bias observed here towards the \code{Internet} category.
Other extensions considered include statistical correlation analyses between libraries categories and \CVE severity, and comparisons of our dataset against Python \PyPI libraries per Topic classifier, e.g.\ on whether the same \FOSS and \CVE filters produce comparable proportions of libraries per category.


\section*{CRediT Authorship Contribution Statement}


\emph{Conceptualization:} FM, CB.
\emph{Methodology:} RP, YF, FM, CB.
\emph{Software Programming:} RP, CB.
\emph{Validation:} CB, FM.
\emph{Investigation:} RP, YF.
\emph{Data Curation:} RP, YF.
\emph{Writing - Original Draft:} CB.
\emph{Writing - Review \& Editing:} RP, YF, FM, CB.
\emph{Visualization:} RP.
\emph{Supervision:} FM.
\emph{Project administration:} CB.
\emph{Funding acquisition:} FM, CB.

\section*{Acknowledgements} 

This work was funded by the EU under GA n.101067199 (\emph{ProSVED}), 101120393 (\emph{Sec4AI4Sec}), and 952647 (\href{https://assuremoss.eu/en/}{\emph{H2020-AssureMOSS}}).

Views and opinions expressed are those of the author(s) only and do not necessarily reflect those of the European Union or The European Research Executive Agency.
Neither the European Union nor the granting authority can be held responsible for them.

\bibliographystyle{elsarticle-harv}  
\bibliography{main}


\begin{appendix}

\section{General guidelines of the protocol}
\label{app:guidelines}

Our protocol provides guidelines, for all actors, along the different steps of the process.
We summarise them below---the full list, including examples and rules of thumb on how to proceed in special cases, can be found in the accompanying dataset of this work \cite{PFMB24}.

\begin{enumerate}[leftmargin=1.5em,parsep=0pt,topsep=.5ex]
\smaller[.5]\slshape
\item
Search and register the sources, which consist of publicly-accessible web links (one link per library per source) stored in a SOURCES file that all assessors can read.
	\begin{enumerate}[leftmargin=1.7em,parsep=0pt,itemsep=0pt,topsep=.5ex]
	\item	The sources to find for each library [\,\ldots] are: Maven repository; library repository; website of the library; wiki page of the library.
	\item	If no (publicly-accessible) web link is found for some of the sources, the corresponding space is left blank.
	\item	For Wiki/Doc pages, links from the same domain than the library's website or GitHub repo are preferred over external sources such as Wikipedia.
	\end{enumerate}
\item
Each assessor reads the sources and chooses a category for the library, that reflects the main purpose of the library—that is saved into a CATEGORIES file that only the arbitrator can read.
	\begin{enumerate}[leftmargin=1.7em,parsep=0pt,itemsep=0pt,topsep=.5ex]
	\item	The available categories to choose from are [\,\ldots]
	\item	All sources must be checked before choosing a category, e.g.\ if the first source identifies the library as ``a database client implementation'', the assessor can choose the ``database'' category only after reading all other sources as well.
	\item	The assessor cannot use sources other than those in the SOURCES file—however, if some technical term is unclear in a source [\,\ldots] the assessor can research its meaning [\,\ldots]
	\item	In the occasions in which two categories seem equally applicable to the library, the assessor can save those two categories, which will have no order of preference.
	\item	Further guidelines are given below [\,\ldots]
	\end{enumerate}
\item
The arbitrator defines the final category for all cases in which there was no unanimous classification by all assessors.
	\begin{enumerate}[leftmargin=1.7em,parsep=0pt,itemsep=0pt,topsep=.5ex]
	\item	The preferred category of a library is the intersection of the categories chosen by all assessors.
	\item	If the intersection is empty or has two categories, the arbitrator must read the sources as in point 2.\ above, and choose one of the categories already pre-selected by the assessors.
	\end{enumerate}
\item
The arbitrator revises the final choice based on NVD evidence.
	\begin{enumerate}[leftmargin=1.7em,parsep=0pt,itemsep=0pt,topsep=.5ex]
	\item	For libraries whose final class is ``Local'', check whether at least one related CVE has an AV=``NETWORK''.
	\item	If so, the arbitrator can choose to change the class of the library to ``Remote network'' [\,\ldots] based on the evidence provided by NVD to assign AV=``NETWORK'' to the relevant CVEs.
	\end{enumerate}
\end{enumerate}

\section{Library (functional) categories}
\label{app:categories}

Our library categories reflect 24 \emph{Topic} classifier from Python Package Index \PyPI---see Topic in \url{https://pypi.org/search/}.
Our protocol expands those categories with the descriptions shown in \Cref{tab:categories} on \cpageref{tab:categories}: this is one of the contributions of this work (see \Cref{sec:intro:contributions}).

%
\begin{table*}
\centering
\caption{24 Topic classifier from \PyPI, used in our protocol as (functional) categories for software libraries}
\label{tab:categories}
\smaller
\colortablepreamble
\begin{tabular}{L{.13\linewidth}L{.7\linewidth}}
	\toprule
	Name
	& Description and examples in our protocol
	\\
	\midrule
	\texttt{Adaptive Technologies}
	& Software used as proxy to interact with other software, e.g.\ to provide better compatibility, or support legacy versions, or add more services to the standard functionality.
	\\
	\texttt{Artistic Software}
	& Code used by digital artists to work, such as illustrators or fonts designers. Do not confuse with \texttt{Multimedia} (see below).
	\\
	\texttt{Database}
	& Libraries whose main purpose is the creation/manipulation/interaction with database systems, including client and server ends.
	\\
	\texttt{Communications}
	& Code used to implement/deploy systems used for communications among humans, such as chat, E-mail, conferencing, IP-telephone, etc.
	\\
	\texttt{Desktop Environment}
	& Similar to \texttt{System} but specifically for the graphical desktop environment, e.g.\ windows manager, system graphical ``themes'', screensavers, etc. (see e.g.\ the Gnome and KDE projects).
	\\
	\texttt{Documentation}
	& Libraries used to generate or process source texttt documentation, such as Doxygen (C/ C++), JavaDoc, Python docstrings, etc.
	\\
	\texttt{Education}
	& Libraries used for educational purposes, such as the ``Snap!'' visual programming language.
	\\
	\texttt{Games/ Entertainment}
	& Software whose main purpose is to contribute to game development, such as a game engine (see e.g.\ the Ogre and Unreal engines), a games platform (Steam, GOG), etc.
	\\
	\texttt{Home Automation}
	& Code intended for domotics, i.e.\ automation of home/buildings appliances, including lighting and sound control, heaters, ventilation, door locking, etc.
	\\
	\texttt{Internet}
	& Software used mainly for remote (aka web) interaction, including client-server protocols, remote filesystems, website development, Internet routing, distributed workflow management, etc. 
	\\
	\texttt{Multimedia}
	& Libraries used for graphics, video, or sound reproduction and manipulation, e.g.\ multimedia players, OBS studio, VLC, Inkscape, alsamixer, OpenAL, etc.
	\\
	\texttt{Office/Business}
	& Libraries used to deploy typical office programs such as word and spreadsheets processors, financial calculators, calendars, etc. Think of Microsoft Office and Outlook.
	\\
	\texttt{Other/Nonlisted Topic}
	& If the library can't even be categorised as \texttt{Utilities} because of its very niche use, it falls into this bucket. E.g.\ a one-use script (that was uploaded as a Maven artifact).
	\\
	\texttt{Printing}
	& Libraries for communication to printers. Consider e.g.\ the CUPs functionality in Linux systems.
	\\
	\texttt{Religion}
	& For the explicit use of religious purposes.
	\\
	\texttt{Scientific/ Engineering}
	& Mostly related with academic research, i.e.\ software used in bleeding edge fields or technologies, such as AI development, physics simulators, theorem provers, medical science, etc.
	\\
	\texttt{Security}
	& Code used to implement/deploy security measures, e.g.\ user authentication, data encryption, secure communication channels, etc. 
	\\
	\texttt{Sociology}
	& Similar to \texttt{Scientific/Engineering} but specifically used for social sciences and in particular sociology. Does not include statistics: that should go in \texttt{Scientific/Engineering}.
	\\
	\texttt{Software Development}
	& Software for the development of more software. Think of IDEs, version control, bug tracking, compilers, QA, testing, etc. ``Documenation'' not included: it has its own category.
	\\
	\texttt{System}
	& Anything used for typical operations in your own local system, e.g.\ file manipulation, resources monitoring, power management and boot, system shell, software package management, etc. 
	\\
	\texttt{Terminals}
	& Libraries for deployment of terminals—not to confuse with ``shells'' which fall under \texttt{System}: terminals are the interface that lets you talk to the shell.
	\\
	\texttt{Text Editors}
	& Libraries used for basic text editing, such as Emacs, Vim, notepad, sublime, etc. Office-specific word processors (like Microsoft Word) don't fall here, but go to \texttt{Office/Business} instead.
	\\
	\texttt{Text Processing}
	& Software for processing (not input) generic text, e.g.\ regular expressions, filtering, markup. Examples are XML-HTML converters, regex filters, JSON serialisers, etc.
	\\
	\texttt{Utilities}
	& Category for ``miscelanea'', i.e.\ anything you could not fit properly in any other category.
	\\
	\bottomrule
  \end{tabular}
\end{table*}

\end{appendix}

\end{document}